\documentstyle[12pt]{article}
\begin{document}
\baselineskip=22pt plus 0.2pt minus 0.2pt
\lineskip=22pt plus 0.2pt minus 0.2pt
\begin{center}
  \Large
ELECTROMAGNETIC DISSOCIATION OF NUCLEI IN HEAVY-ION COLLISIONS\\
\vspace{0.35in}
by\\
\vspace{0.35in}
  \large
C.\ J.\ Benesh \\
Theoretical Division\\
Los Alamos National Laboratory\\
Los Alamos, NM 87545\\
\vspace{0.5in}
J.\ L.\ Friar\\
Theoretical Division\\
Los Alamos National Laboratory\\
Los Alamos, NM 87545\\
\vspace{0.15in}
and\\
\vspace{0.15in}Institut f\"{u}r Kernphysik\\
Johannes Gutenberg-Universit\"{a}t\\
D-6500 Mainz, Germany\\
\end{center}
\baselineskip=20pt plus 0.2pt minus 0.2pt

\begin{abstract}
	Large discrepancies have been observed between measured
Electromagnetic Dissociation(ED) cross sections and the predictions
of the semiclassical Weiz\"acker-Williams-Fermi(WWF) method. In this paper,
the validity of the semiclassical approximation is examined.
The total cross section for electromagnetic
excitation of a nuclear target by a spinless projectile is calculated
in first Born approximation, neglecting recoil. The final result
is expressed in terms of correlation functions and convoluted densities
in configuration space. The result agrees with the WWF approximation to
leading order(unretarded electric dipole approximation), but the method
allows an analytic evaluation of the cutoff, which is determined by
the details of the electric dipole transition charge density.
Using the Goldhaber-Teller model of that density, and uniform
charge densities for both projectile and target, the cutoff is determined
for the total cross section in the nonrelativistic limit, and found to be
smaller than values currently used for ED calculations.

In addition, cross sections are calculated using a phenomenological
momentum space cutoff designed to model final state interactions. For
moderate projectile energies, the calculated ED cross section
is found to be smaller than the semiclassical result, in qualitative
agreement with experiment.
\end{abstract}
\pagebreak
\baselineskip= 22pt plus 0.2pt minus 0.2pt
\section{Introduction}
	The availability of relativistic heavy ion beams has opened a
new avenue for the study of electromagnetic  excitations of nuclei.
Cross sections are
enhanced both by the charge of the projectile ions, and
by the relativistic contraction of the projectile's electric field into
a sharp pulse of radiation at high energies. Experiments range from
single and double nucleon-removal reactions\cite{2}, to the
study of ``halo'' nuclei using radioactive beams\cite{1},
to the possibility of
multi-phonon excitations of collective nuclear states\cite{3}. Aside from
their intrinsic interest, electromagnetic excitation processes in peripheral
collisions will also be important at RHIC\cite{3.5}.

There is,
therefore, a tremendous incentive to develop an understanding of the
physics involved in these processes. Such an understanding has two facets.
On the one hand, one must be able to calculate the detailed structure of
the target nucleus in order to calculate its response to a particular probe,
such as the electromagnetic interaction.
Fortunately, this has been one of the central topics of nuclear physics
for many years, and an extensive literature exists on the subject\cite{4}.
On the other hand, it is also necessary to understand the process by which
the projectile excites the target. With few exceptions to date\cite{5},
this aspect of the problem has been dealt with by using the venerable
Weiz\"acker--Williams--Fermi(WWF) method of virtual quanta\cite{2a}, and its
generalization to arbitrary multipoles\cite{7a}. This approach is based on
the observation, due to Fermi, that the electromagnetic fields of a
point charge, when boosted to high energy, are transverse to the direction
of the charge's motion. The supposition is that one can calculate
cross sections by replacing the projectile by an equivalent pulse
of electromagnetic radiation. The cross-section for a given reaction is then
\begin{equation}
\sigma_{\rm WW}=\int d\omega \, n(\omega)\, \sigma_{\gamma}(\omega),
\end{equation}
where $\sigma_\gamma(\omega)$ is the cross section for the same reaction
induced
by real photons, and $n(\omega)$ is the number of  photons of energy
$\omega$ in the pulse of equivalent radiation. For dipole transitions,
$n(\omega)$ can be determined by calculating the intensity of the projectile
fields as a function of frequency, integrated over impact parameters\cite{6}.
Assuming the projectile's trajectory is a straight line, the result is
\begin{eqnarray}
n_{E1}(\omega)&=&{2Z_p^2\alpha\over \pi\beta^2\omega}
(xK_0(x)K_1(x)-{\beta^2
\over 2}x^2[K_1^2(x)-K_0^2(x)])\nonumber\\
&\sim &{2Z_p^2\alpha\over \pi\beta^2\omega}
(\log({1.123\over x})-{\beta^2\over 2})\quad \quad {\rm as}\  \gamma
\rightarrow \infty \ ,
\end{eqnarray}
where $K_{0(1)}$ is a Bessel function of imaginary argument,
$Z_p$ is the projectile charge, $\beta$ its speed,
$x={\omega b_{min}\over\gamma\beta}$, and $b_{min}$ is an impact parameter
cutoff required to get a finite result.

	In some respects, the Weiz\"acker--Williams--Fermi(WWF) method is very
well suited to the problem of electromagnetic dissociation by heavy ions.
Cross sections may be calculated either
using a model, or, as is often done, by direct appeal to measured
photodisintegration cross sections.
In the latter case one obtains a cross section,
at least at high energy\cite{7}, that is independent of any model of
the target's structure. Furthermore, the equivalent photon number,
$n(\omega)$, can be obtained by a relatively straightforward classical
calculation. Nonetheless, the WWF method is not completely satisfactory.
Theoretically, there is no systematic procedure for evaluating corrections
to the semiclassical result, and consequently no way of gauging the
reliability of the approach. In addition, there is the troublesome problem
of choosing the minimum impact parameter. Since $b_{min}$ is not fixed by the
WWF procedure itself, a number of choices have been proposed\cite{8}, based
on phenomenological considerations. On the experimental side, large
discrepancies from the naive WWF predictions have been observed
in target fragmentation experiments\cite{9}, leading some authors to question
the validity of first order perturbation theory for heavy
projectiles\cite{9.1}.

	The aim of this paper is to go beyond the semiclassical WWF
method and calculate the quantum mechanical cross section for
electromagnetic excitation of nuclei in heavy ion collisions. In the next
section, we derive an expression, originally used in high energy physics
\cite{10}, for the unpolarized cross-section in
first Born approximation, neglecting recoil effects for both the
projectile and target nuclei. We demonstrate that the cross section
does approach the WWF approximation in the limit of large projectile
energy, and in an appendix we derive an expression for the cutoff
parameter,$b_{min}$, for dipole transitions, using a simple model for
the transition matrix
elements. In the following section, we use the same simple model to
compare our results for the full cross section to those of the WWF
approximation as a function of projectile energy, transition multipolarity,
and transition frequency. Finally, using a phenomenological cutoff designed
to model final state interaction effects, we compare the results of this simple
model with measured single-neutron removal cross sections.

\section{Quantum Excitation Cross Section}

	We begin this section with a discussion of the kinematics
of the nuclear excitation process.  The projectile
nucleus, with mass $M_1$ and momentum $P^\mu_i=(E_i,\vec P_i)$, scatters
from the target by exchanging a virtual photon of momentum $q^\mu$.
In the process, depicted in figure 1, the target of mass $M_2$ is excited from
its ground
state to an excited state of mass $M_2+\omega$. For high
$Z_p$ projectiles, elastic scattering of the projectile($\propto Z_p^2$)
will dominate over inelastic processes($\propto Z_p$), so that in the
following we shall
assume that the projectile remains unexcited in the final state. This is in
accord with the semiclassical picture, where the projectile remains in its
ground state, following a
straight line trajectory throughout the collision. Kinematically, the
elastic scattering of the projectile translates into a condition on the
four momentum transfer $q$,
\begin{equation}
q^2=2P_i\cdot q\qquad\rightarrow\qquad q^2=2E_i(q_0-\vec\beta\cdot\vec q),
\end{equation}
where $\vec\beta=\vec P_i/E_i$ is the projectile velocity. For nuclear
transitions, the momentum and energy transfers are on the order of tens
of MeV, while the projectile and target masses are on the order of tens
of GeV. As a result, $q^2/2E_i$ is negligible, and we immediately obtain
the minimum momentum transfer
\begin{equation}
\vert{\vec {\,q}\,}\vert_{min}=q_0/\beta\qquad\rightarrow\qquad
q_{min}^2=-{q_0^2\over \gamma^2\beta^2},
\end{equation}
where $\gamma={1\over\sqrt{1-\beta^2}}$. In exactly the same fashion,
we derive a relation between the energy transfer and the target
excitation energy,
\begin{equation}
q_0=\omega+{\cal O}(\omega^2/M_2).
\end{equation}
Even at this stage of the calculation, we see a similarity with
the semiclassical calculation at high energy. For large projectile
energies, the minimum momentum transfer goes to zero, so that it is reasonable
to expect that the cross section will be dominated by the pole in the
photon propagator at $q^2=0$,
where  the interaction cross section will be well approximated by the cross
section for real photons.

	Having treated the kinematics, we now turn to the calculation of the
cross section. The method we use is a straightforward evaluation of the
Feynman diagram of figure 1, borrowing heavily from the theory of
relativistic electron scattering\cite{1a} in order to relate the target and
projectile form factors to the appropriate nuclear transition matrix elements.
For a 0$^+$ projectile (the generalization to nonzero spin is trivial),
the spin averaged cross section for an arbitrary transition multipole is
given by
\begin{eqnarray}
\sigma_{\rm WW}&=-{(4\pi Z_p\alpha)^2\over ((P_i\cdot K_i)^2-M_1^2M_2^2)^{1/2}}
&\int \,{d^4q\over (2\pi)^4} {d^3P_f\over 2E_f (2\pi)^3}
 (2\pi)^4\delta^4(P_f-P_i+q)\nonumber\\
& &\times F_{p}^2(q^2)
P_i^\mu P_i^\nu W^T_{\mu\nu}(q^2,q\cdot K_i),
\end{eqnarray}
where the matrix element of the projectile current is given by
$Z_p\, F_{p}(q^2)\,(P_{i\mu}+{q_\mu\over 2})$,
and the spin-averaged target structure function is given by
\begin{eqnarray}
W^T_{\mu\nu}(q^2,q\cdot K_i)&=&\int {d^3K_f\over (2E_{K_f})(2\pi)^3}
(2\pi^4)\delta^4(K_f-K_i-q){1\over(2J_i+1)}\nonumber\\
& &\sum_{mm^\prime} \langle K_i,m\vert J_\mu(0)\vert K_f,m^\prime\rangle
\langle K_f,m^\prime\vert J_\nu(0)\vert K_i,m\rangle \\
&=& W_1(q^2,q\cdot K_i)\big{(}g_{\mu\nu}-{q_\mu q_\nu\over q^2}\big{)}
\nonumber\\
& &\qquad +W_2(q^2,q\cdot K_i)\big{(}K_{i\mu}-{q\cdot K_iq_\mu\over q^2}\big{)}
\big{(}K_{i\nu}-{q\cdot K_iq_\nu\over q^2}\big{)},
\end{eqnarray}
where $m(m^\prime)$ denote the azimuthal angular momentum quantum
numbers of the initial(final) state, $(E_{K_f},\vec K_f)$ is the
final state four
momentum, and  $W_1(q^2,q\cdot K_i)$ and $W_2(q^2,q\cdot K_i)$ are
Lorentz invariant target structure functions. For what follows, it is
critical to realize that while the structure functions $ W_1(q^2,q\cdot K_i)$
and $ W_2(q^2,q\cdot K_i)$ are different for different transition multipoles,
the tensor form of $W^T_{\mu\nu}$ is determined by current conservation
and parity, and thus independent of both the multipolarity of the transition
and spin of the target. Using the elastic scattering condition for the
projectile, and keeping only the leading terms in inverse powers of the
projectile and target masses, we get
\begin{eqnarray}
\sigma_{\rm WW}&=-{(4\pi Z_p\alpha)^2\over ((P_i\cdot K_i)^2-M_1^2M_2^2)^{1/2}}
&\int\,{d^4q\over (2\pi)^4} {d^3P_f\over 2E_f (2\pi)^3}
 \,(2\pi)^4\delta^4(P_f-P_i+q) \,F_{p}^2(q^2) \nonumber\\
& &\times\Big{[}M_1^2W_1(q^2,q\cdot K_i)+(P_i\cdot K_i)^2W_2(q^2,q\cdot K_i)
\Big{]}.
\end{eqnarray}
At this point, it is useful to reexpress the two invariant
structure functions in terms nuclear matrix elements.
In particular, we shall choose the two rotationally invariant combinations
$\langle J_0(0)J_0(0)\rangle \equiv \langle \rho(0)\rho(0)\rangle$
and $\langle \vec J(0)\cdot \vec J(0)\rangle$,
where the brackets are introduced as a convenient shorthand for the
spin sums and matrix elements of equation 7. Hence, in the target rest frame,
\begin{eqnarray}
W_1(q^2,q\cdot K_i)&=&\int {d^3K_f\over (2E_{K_f})(2\pi)^3}
(2\pi)^4)\delta^4(K_f-K_i-q)\nonumber\\
& &\times{-\vec q^2\langle{\vec J(0)\vec J(0)}\rangle +q_0^2\langle
\rho(0)\rho(0)\rangle
\over 2\vec q^2}\\
W_2(q^2,q\cdot K_i)&=&\int {d^3K_f\over (2E_{K_f})(2\pi)^3}
(2\pi)^4\delta^4(K_f-K_i-q){q^2\over M_2^2}\nonumber\\
& &\times{(2q^2+q_0^2)\langle \rho(0)\rho(0)\rangle
-\vec q^2\langle {\vec J(0)\vec J(0)}\rangle \over \vec q^4}
\end{eqnarray}
	For analytical purposes, it is convenient to have expressions in
configuration space. Replacing the momentum delta functions by an integral
representation, and using translational invariance, we obtain
\begin{eqnarray}
W_1(q^2,q\cdot K_i)&=&\int {d^3K_f\over (2E_{K_f})(2\pi)^3}
(2\pi)\delta(E_{K_f}-M_2-q^0)\nonumber\\
& &\times\int d^3z\, e^{i\vec q\vec z}\,
{-\vec q^2\langle {\vec J(\vec z)}{\vec J(0)}\rangle
+q_0^2\langle \rho(\vec z)\rho(0)\rangle \over 2\vec q^2}\\
W_2(q^2,q\cdot K_i)&=&\int {d^3K_f\over (2E_{K_f})(2\pi)^3}
(2\pi)\delta(E_{K_f}-M_2-q^0){q^2\over M_2^2}\nonumber\\
& &\times\int d^3z\, e^{i\vec q\vec z}\,{(2q^2+q_0^2)\langle \rho(\vec z)
\rho(0)\rangle -\vec q^2\langle \vec J(\vec z)\vec J(0)\rangle \over \vec q^4}
\end{eqnarray}
	The projectile form factor is related to the projectile {\it
rest frame} charge density via
\begin{equation}
F(q^2)=\int\,d^3x e^{i\vec q\,^\prime\cdot \vec x} \rho_{p}(\vec x),
\end{equation}
where $\vec q\,^\prime = \vec q-{(\gamma -1)\over\gamma\beta^2}\vec\beta\cdot
\vec q\vec\beta$ is the three momentum transfer in the projectile rest frame.

	Since the matrix elements appearing here are rotational scalars,
we may replace the complex exponential by its angular average with impunity.
The result for the cross section then becomes, after performing the integrals
over the three momentum transfer,
\begin{eqnarray}
\sigma_{\rm WW}&={2(Z_p\alpha)^2\over \beta^2}&\int {dq^0\over (q^0)^2}
\int {d^3K_f\over (2\pi)^3
2E_{K_f}} (2\pi)\delta(E_{K_f}-M_2-q^0)\nonumber\\
& &\times\int d^3z \big{[}
\langle \rho_c(\vec z)\rho_c(0)-\vec J_c(\vec z)\vec J_c(0)\rangle
({1\over \gamma^2}I_1(q_0z,\beta)+I_2(q_0z,\beta))
\nonumber\\
& &+\langle \rho_c(\vec z)\rho_c(0)\rangle
({1\over\gamma^2}I_2(q_0z,\beta)+I_3(q_0z,\beta))\big{]}\ ,
\end{eqnarray}
where
\begin{eqnarray}
I_1(y,\beta)&=&\int_{1\over \beta}^\infty q\,dq {j_0(qy)\over
(1-q^2)^2}\nonumber\\
&=& {1\over 4y}\big{[}(Si((1/\beta-1)y)-Si((1/\beta+1)y)(\cos y+y\sin y)
\nonumber\\
& &+(Ci((1/\beta+1)y)+Ci((1/\beta-1)y))(\sin y-y\cos
y)+2\beta\gamma^2\sin(y/\beta)\big{]},\nonumber\\
& &\\
I_2(y,\beta)&=&\int_{1\over \beta}^\infty q\,dq {j_0(qy)\over
q^2(1-q^2)}\nonumber\\
&=&{1\over y}\big{[}{\textstyle{1 \over 2}} \cos
y(Si((1/\beta-1)y)-Si((1/\beta+1)y))\nonumber\\
& &+{\textstyle{1 \over 2}} \sin y(Ci((1/\beta+1)y)+Ci((1/\beta-1)y))
+\beta \sin(y/\beta)-yCi(y/\beta)
\big{]},\nonumber\\
& &\\
I_3(y,\beta)&=&3\int_{1\over\beta}^\infty {q\,dq\over q^4} j_0(qy)\nonumber\\
&=&(\beta^2-y^2/2)j_0(y/\beta)+{\beta^2 \over 2} \cos(y/\beta)+{y^2 \over 2}
\,Ci(y/\beta),
\nonumber\\
\end{eqnarray}
with $Si(y)$ and $Ci(y)$ the sine and cosine integral functions,
respectively, and
\begin{eqnarray}
\rho_c(\vec z)&=&{\gamma\over Z_p}\int d^3x \rho_{p}(\vec x-{\gamma^2\over
\gamma\beta^2}\vec\beta\cdot\vec x\vec\beta)\rho(\vec z-\vec x),\nonumber\\
\vec J_c(\vec z)&=&{\gamma\over Z_p}\int d^3x \vec J_{p}(\vec x-{\gamma^2\over
\gamma\beta^2}\vec\beta\cdot\vec x\vec\beta)\rho(\vec z-\vec x).
\end{eqnarray}

	At this juncture, the expression bears little resemblance to
the compact semi-classical result in equations 1 and 2. There is no
simple factorization of the integrand into a flux factor multiplying
the photo-cross section. The integrand is not even a function of the
same variable, $q^0/\gamma\beta$, as the expressions in equation 1.
Asymptotically, the cross section of equation 14 falls off like
a power for large $q^0$, in contrast to the exponential decrease
dictated by the Bessel functions of the semiclassical result. In light of
this, it seems quite unlikely that
the semiclassical expression will yield a good approximation for the
cross section over a large energy range.

On the other hand, the correspondence
principle tells us that the two expressions must agree in the classical limit.
For the problem at hand, the classical regime occurs for large projectile
energies, where the straight line trajectory of the projectile represents a
good approximation to the classical trajectory. In that limit
($\gamma\rightarrow\infty$), the momentum integrals become
\begin{eqnarray}
I_1(y,\beta)&\approx & {\gamma^2 \over 2} j_0(y)+ {\cal O} (1), \\
I_2(y,\beta)&\approx & -j_0(y)\log (\gamma)+ {\cal O} (1),\\
I_3(y,\beta)&\approx & {\cal O}(1),
\end{eqnarray}
and the cross section is given by
\vfill\eject
\begin{eqnarray}
\sigma_{\rm WW}&=&{(Z_p\alpha)^2\over\beta^2}\int {dq^0\over (q^0)^2}
\int {d^3K_f\over (2\pi)^3
2E_{K_f}} (2\pi)\delta(E_{K_f}-M_2-q^0)\nonumber\\
& &\times\int\, d^3z \Big{[}\langle \rho_c(\vec z)\rho_c(0)-\vec J_c(\vec z)
\cdot \vec J_c(0)\rangle
\, j_0(q^0z)\,
\big{(} 1-\log(\gamma^2)\big{)} +{\cal O}(1)\Big{]}.
\end{eqnarray}
Using methods identical to those just described, the cross section for
real photons can be written as,
\begin{eqnarray}
\sigma_\gamma(q^0)&=&-{\pi\alpha\over q^0}\int {d^3K_f\over (2\pi)^3
2E_{K_f}} (2\pi)\delta(E_{K_f}-M_2-q^0)\nonumber\\
& &\,\,\,\times\int d^3z \, j_0(q^0z)
\langle \rho(\vec z)\rho(0)-J(\vec z)J(0)\rangle .
\end{eqnarray}
{}From which it follows that
\begin{eqnarray}
\sigma_{\rm WW}&=&{2(Z_p\alpha)^2\over \beta^2}\int {dq^0\over (q^0)^2}\Big{[}
\big{(}\log({2\gamma\over q^0R_\ell})-\gamma_E-{1\over 2}\big{)}{q^0\sigma
_\gamma(q^0)\over \pi\alpha}\nonumber\\
& &+{\cal O}(1/\gamma^2)\Big{]},
\end{eqnarray}
where all the terms of order $(1/\gamma)^0$ have been absorbed into
the cutoff parameter $R_\ell$.

The first term in this expression is just the high energy limit of
the semi-classical cross section, with no cutoff parameter.
The cutoff parameter, $R_\ell$, represents the leading order
corrections due to Coulomb
and longitudinally polarized virtual photons, as well as off shell
corrections to the  photonuclear cross section. In the limit of
large $\gamma$, the precise value of $log(q^0R_\ell)$ will be
negligible compared to $log(\gamma)$, and the semiclassical
result is recovered.

	In the low frequency limit, the off-shell corrections to the
photonuclear cross section vanish, and $R_\ell$ is determined solely by the
projectile and target transition densities. In Appendix B, we exploit
this fact to calculate $R_1$,
using the Goldhaber-Teller model\cite{5a} for the transition densities.
Accurate determination of this parameter, which governs the {\it total}
excitation cross section, will be relevant for estimating the
large background due to electromagnetic processes at RHIC, where the
semiclassical limit should be realized.

\section{Results}

	In this section, we compare results both with the
semiclassical calculations using the Weiz\"acker-Williams method,
and,ultimately, with data from single neutron removal experiments.
To do this, we are required to
assume a model for the transition matrix elements appearing in
eq. 14. For simplicity's sake, we again choose the Goldhaber-Teller(GT)
\cite{5a} model to describe the transition densities, and assume that
the projectile and target are described by uniform density spheres.
A brief sketch of
the GT model, as applied to dipole excitations, may be found in the appendix.
For general multipoles, a more detailed description of the model, including
electromagnetic transition matrix elements, may be found in ref. 15.

	In order to effect a comparison with the
semiclassical calculation for arbitrary $\gamma$, it is useful
to decompose the total cross section into multipoles.
Since the magnetic multipole contributions to the cross section are small,
we shall restrict our attention to electric multipoles larger than zero.
The decomposition may be accomplished by inserting a factor
\begin{equation}
1={1\over V}\int d^3x\,d^3x^\prime \delta^3(\vec z-\vec x+\vec x^\prime),
\nonumber
\end{equation}
into eq. 14, and by using the Bessel function identity
\begin{equation}
j_0(q\vert \vec x-\vec x^\prime\vert)={1\over 4\pi}\sum_{\ell m}
j_\ell (qx)j_\ell (qx^\prime)Y_{\ell m}(\hat x)Y_{\ell m}^*(\hat x^\prime),
\end{equation}
to do the angular integrals. The result is
\begin{eqnarray}
\sigma_{\rm WW}&=&{2(Z_p\alpha)^2\over \beta^2}\sum_\ell \int\, dq^0
\delta(q^0-\omega_\ell)
\int_{q_0/\beta}^\infty q\,dq F_p^2(q^\prime)\nonumber\\
& & \Big{[} {1\over\gamma^2}
{(F^\ell_\rho(q)-F^\ell_J(q))
\over (q_0^2-q^2)^2}+{(F^\ell_\rho(q)-F^\ell_J(q))\over (q_0^2-q^2)q^2}
+{F^\ell_\rho(q)\over q^4}+{1\over\gamma^2}
{F^\ell_\rho(q)\over (q_0^2-q^2)q^2}\Big{]},\nonumber\\
\end{eqnarray}
where, for $\ell >0$, $F_\rho^{\ell}(q)=\ell C_\ell j_\ell^2(qR_t)$, and
$F_J^\ell(q)={(2\ell+1)\over\ell} {q_0^2\over q^2} F_\rho^\ell(q)$, with
$R_t$ the radius of the target nucleus and $C_\ell$ a constant chosen
such that the photo-cross section satisfies the energy weighted sum rule
for multipole $\ell$, and $F_p(q^\prime)=3j_1(q^\prime R_p)/q^\prime R_p$ is
the elastic form factor of the projectile, with $R_p$ the projectile radius.

	For comparison, we also calculate the semiclassical cross section,
using a prescription from reference 11 for the minimum impact parameter,
\begin{equation}
b_{min}=1.34\Big{[}A_p^{1/3}+A_t^{1/3}-0.75(A_p^{-1/3}+A_t^{-1/3})\Big{]},
\end{equation}
appropriate for single and double nucleon removal experiments.
The ratio of the quantum and semiclassical cross sections for
$^{12}$C and $^{197}$Au projectiles incident on a $^{197}$Au target are
shown,as a function of projectile energy, in figures 2 and 3, respectively,
for both E1 and E2 transitions. We assume that the E1(E2) transition is to a
sharp, isovector(isoscalar) giant resonance state at 13.8(11.0) MeV, and that
the energy weighted sum rule is saturated. We further assume that the giant
resonance state decays exclusively via one neutron emission.
For high projectile energies,
we find that the quantum E1 cross section is enhanced by about 10 per cent
relative to the
classical result. This enhancement agrees with the results of the appendix,
where it was shown that the quantum mechanical cutoff is smaller than
that of eq. 29, resulting in a larger cross section. More surprising is
the
enhancement of the E2 cross section, which remains a factor of two larger than
the classical result at $\gamma$=100. Also noteworthy is the huge enhancement
of both the quantum E1 and E2 cross sections at small projectile energies. For
$\gamma >$5, nearly all of the difference in the E1 cross section may be
reabsorbed into a redefinition of $b_{min}$ as described in the appendix,
so we conclude that the WWF method is a good approximation for the
total E1 cross section for energies of 5 GeV/nucleon and higher, provided
that the minimum impact parameter is properly chosen. For lower
energies, much of the difference can be eliminated by altering $b_{min}$,
but discrepancies as large as 20 per cent persist at very low projectile
energies.

	In figures 4 and 5, the ratio of the quantum and semiclassical
cross sections is shown as a function of energy for $\gamma$=2,
assuming that the momentum dependence of the transition matrix elements
does not vary with transition energy. As $\omega$ goes to zero,
the E1 cross section approaches the semiclassical result.
This is expected, as in this limit the argument of the logarithm is large
in both the semiclassical expression and the high energy limit of the quantum
calculation. As a result, the semiclassical piece of the quantum cross-section
dominates, and the two limits are the same. For large $\omega$, both the E1
and E2 ratios are enhanced, and the degree of enhancement is quite sensitive
to the behavior of the transition form factors. For small $\omega$, the
quantum E2 cross section is roughly four times the classical result,
indicating that the large $\gamma$ and small $\omega$ limits are not
equivalent for higher multipoles.

	In order to make a meaningful comparison with experiment, we must
account for the final state strong interactions between the target and
projectile.
The choice of $b_{min}$ in eq. 29 is designed to do exactly this for
the semi-classical problem\cite{8}. Briefly, $b_{min}$ is chosen such
that the mean number of nucleon-nucleon collisions, calculated using
the Glauber approach , at impact parameter $b=b_{min}$ is exactly one. For
$b>b_{min}$, the probability that a second nucleon will be knocked out
of the target by the strong interaction drops rapidly to zero. For
$b<b_{min}$, the mean number of nucleon-nucleon collisions rises very
rapidly, and the probability of at least one additional nucleon getting
out of  the target goes rapidly to one. Hence, we obtain the usual
semiclassical picture, where the equivalent photon number is calculated
by integrating the field intensity from all trajectories with $b>b_{min}$.

	The same idea can be applied to the quantum picture by artificially
including a maximum value, $q_{max}\approx 1/b_{min}$, for the transverse
momentum transfer in the collision. The effect of this cutoff is to change
the upper limit of the $q$ integration in eq. 28
to $\sqrt{q_0^2/\beta^2+1/b_{min}^2}$
instead of $\infty$. In figures 2-5, the ratio of the new quantum and
semiclassical cross sections is shown as a function of the projectile energy
and transition frequency for the same projectile target combinations
used previously. The same general trends hold; the quantum cross section is
enhanced for small projectile energies, and approaches the semiclassical limit
as $\gamma$ gets large.(This latter fact holds because the cross section
is dominated by the photon pole as $\gamma\rightarrow\infty$, so that the
precise value of the high $q$ cutoff becomes irrelevant.) Not unexpectedly,
the size of the quantum cross section is smaller than in the case of no
cutoff, with the result that the small $\gamma$ enhancement of the cross
section is less pronounced, and, for large $\gamma$,
the semiclassical limit is approached from
below, rather than above as before. Similar conclusions may be drawn
regarding the transition frequency dependence of the cross section. As before,
the small $\omega$ and large $\gamma$ limits of the E1 cross section
approach the classical result, while the two limits differ markedly for
the quadrupole cross section. Perhaps the most noteworthy feature of the
$\omega$ dependence is that the strong dependence of the cross section ratio
on the transition form factors has been largely eliminated.

	Of particular phenomenological interest is the
20-25 per cent suppression of the quantum E1 cross section relative to the
semiclassical in the region $\gamma=2-3$. This suppression is at precisely
the right location and magnitude to explain the discrepancy between the
WWF approximation and recent single neutron removal data for $^{238}$U
on $^{197}$Au\cite{9}. In table 1, we compare the results of the our
calculation and the WWF calculation  with experimental data from references 2
and 12. For the heaviest projectiles, the agreement between the quantum theory
and experiment is much improved over the WWF results. For Fe and Ar, the
quantum theory does about as well as the semiclassical, and for the lightest
two projectiles, the semiclassical theory does better.

This level of
agreement with the quantum theory is in fact quite heartening.
To begin with, the ``experimental'' numbers listed in table 1
are not raw data, but rather the difference between the raw data and an
estimate of the strong interaction contribution to the single neutron
removal cross section. As noted in reference 11, estimates of the strong
interaction contribution to the cross section are model dependent, so that
an additional systematic uncertainty of at least 20 mb should be added
to all the results listed in table 1. Clearly, for the heavy projectiles,
this additional uncertainty is not important, as the extracted electromagnetic
cross sections are quite large by comparison. For the lightest projectiles,
however, the additional uncertainty is comparable to the extracted ED
cross section, and is quite likely responsible for the observed discrepancies.

\section{Conclusions}

	Using a very simple model of nuclear structure, we have
calculated the electromagnetic excitation cross section for
nuclear collisions in first Born approximation,
neglecting final state interactions, and have found significant
differences from results of the semiclassical WWF method. For E1 transitions,
we find that the impact parameter cutoff required for the semiclassical
calculation to agree with the high energy limit of the quantum cross section
is significantly smaller than the phenomenological cutoffs used to
analyze experiments. At energies less than a few GeV/nucleon, the cross
section is enhanced over the semiclassical result by as much as 20
per cent when the smaller cutoff is used. For E2 transitions, we find
the cross section is significantly enhanced even at RHIC energies, and,
unlike the E1 case, that the limit of small transition energy and
large projectile energy are not the same.

	When final state interactions are included via a phenomenological
cutoff, we find that the E1 cross section is greatly enhanced over the
analogous WWF calculation for low projectile energies, while the cross
section is suppressed at higher energies. The E2 cross section is suppressed
by as much as factor of three for all but the lowest projectile energies.
The low energy enhancement is relevant for electromagnetic dissociation
studies of the low energy, pygmy resonances in neutron rich nuclei such as
$^{11}Li$, and the suppression at higher energies resolves the conflict
between single neutron
removal experiments and the semiclassical theory for all but the lightest
projectiles, where the data is very sensitive to systematic errors in the
separation of the nuclear and electromagnetic contributions to the cross
section.

	While the agreement with data is extremely satisfactory given
the simplicity of the Goldhaber-Teller model, a number of interesting
questions remain. To begin, the calculation should be redone using
a realistic model for the projectile and target densities in order to
improve the quantitative description of the data. In addition, the effect
of additional photon exchanges should be studied, both to understand the
effect of the repulsive coulomb potential on the scattering process, and
to study the question of multi-phonon excitations in the target nucleus.
Finally, the process where both the target and projectile are excited,
which should be non-negligible in target fragmentation experiments
with light projectiles, should be calculated.
\pagebreak

\begin{center}
    \large
{\bf Acknowledgements}
\end{center}

The work of C.\ J.\ B.\ and J.\ L.\ F.\ was performed under the
auspices of the
U.\ S.\ Department of Energy. One of us (J.\ L.\ F.\ ) would like to
thank the
Alexander von Humboldt-Stiftung for support.

\pagebreak

\noindent{\bf Appendix A}

In this appendix we sketch the cross section derivation for target excitation
in a more familiar form, which highlights the nonrelativistic nature of
the nuclear physics. It also allows us to
point out our approximations (explicit or implicit) and to emphasize the
role of gauge invariance, which puts the results into a form where Siegert's
theorem can be applied immediately.  We assume the following:
\begin{enumerate}
\item{} First Born approximation in the fine-structure constant,
$\alpha$; this is required for tractability.

\item{}  No target recoil, which presupposes that momentum transfers
are very small compared to the target mass, $m_t$; this greatly simplifies
the kinematics.

\item{}  The maximum momentum transfer can be replaced by infinity, for
ease of performing integrals; in practice, small momentum transfers dominate.

\item{}  A nonrelativistic description of the target nucleus,
and of the \underline{internal} structure of the projectile nucleus, is
sufficient; photonuclear physics is almost entirely nonrelativistic,
and most of what we know is based on this (successful) description.

\item{}  Both target and projectile are spinless and only the former
is excited; neither restriction is essential, but will simplify the
derivation.

\item{}  All purely hadronic contributions to the excitation
amplitude can be ignored. Elastic Coulomb scattering is infinite and
will also be ignored.
\end{enumerate}

The current of the elastically scattered projectile is conserved and is
given by $F_p(q^2)(P_f + P_i)^{\mu} / \sqrt{4E_f E_i}$, where the projectile
charge distribution is present through its Fourier transform, the projectile
form factor, $F_p(q^2)$.  The latter is a function of the (squared)
four-momentum transfer, $q^{\mu} = P_i - P_f$.  The distinction between this
quantity and the three-momentum transfer is a relativistic correction (and a
recoil correction in the projectile rest frame).  We consequently replace
$F_p(q^2)$ by $F_p(\vec{q}^2) \equiv \int d^3x\, \rho_p(x) \, e^{i\vec{q}\cdot
\vec{x}}$.

The target current is denoted by $\hat{J}^{\mu}(\vec{q}) =
(\hat{\rho}(\vec{q})), \vec{\hat{J}}(\vec{q}))$ and is conserved.  That is,
$q_{\mu} \hat{J}^{\mu}(\vec{q}) \equiv 0$.  Rewriting the kinematical factors
in
the projectile current, $(P_f + P_i)^{\mu}$, as $2P_i^{\mu}-
q^{\mu}$, we can use current conservation and drop the $q^{\mu}$ factor.
Thus, the current-current matrix element is given by
$(E_i / E_f)^{\frac{1}{2}}$ $(\rho_{N0}(\vec{q}) - \vec{\beta} \cdot
\vec{J}_{N0}(\vec{q}))$, where $\vec{\beta} = \vec{P}_i / E_i$ is the
projectile
velocity and $J_{N0}^{\mu}(\vec{q}) \equiv <N|\hat{J}^{\mu}(\vec{q})|0>$. The
energy difference $q_0$ of the states labelled $N$ and $0$ is denoted by
$\omega_N$.

We wish to calculate the total cross section, which means that we must
integrate
over all of phase space.  Ignoring recoil (which means that $P^{\mu}_f$ is
independent of the scattering angle, $\theta$), we have for fixed energy
transfer:
$\vec{q}^2 =  (\vec{P}_f - \vec{P}_i)^2 = P_f^2 + P_i^2 - 2P_f P_i cos \theta$,
or
$d\vec{q}^2 = 2P_f P_i sin \theta d \theta = d \Omega(2P_f P_i /2 \pi)$, since
the azimuthal  dependence is trivial.  All of the phase-space integrals (but
one) can now be performed and we obtain:
$$
\sigma_{\rm WW} = \frac{4 \pi Z_p^2 \alpha^2}{\beta^2} \sum_{N\neq 0}
\int^{\infty}_{\vec{q}^2_{min}} \frac{d \vec{q}^2 F^2_p (\vec{q}^2)}
{(\vec{q}^2 - \omega^2_N)^2} | \rho_{N0} (\vec{q}) - \vec{\beta} \cdot
\vec{J}_{N0}(\vec{q})|^2. \eqno({\rm A}1)
$$
This deceptively simple form has many aspects of complexity.  The integral
should be dominated by the pole in the photon propagator and the small values
of
$\vec{q}^2_{min} = \omega^2_N /\beta^2$ and $q^2_{min} = -
\omega^2_N /\beta^2\gamma^2$.  This further implies that electric dipole
processes should dominate, because they are largest for small $\vec{q}^2$.
This multipole should consequently not be very sensitive to $F_p$, unlike
higher ones which are certain to be.  In addition, we can make use of the
conserved current to transform to Coulomb gauge.  This eliminates the
(redundant) longitudinal component of the current, and expresses the result
in terms of transverse-current matrix elements (which determine
photoabsorption)
and purely Coulombic excitation.  We also impose the no-recoil
approximation in the form $P_i \cdot q/E_i =
q_o - \vec{\beta} \cdot \vec{q} = q^2/2E_i \sim 0 $, so that
we can replace $\vec{\beta} \cdot \vec{q}$ by $\omega_N (=q_0)$ where needed.
We also
impose current conservation: $\vec{q} \cdot \vec{J}_{N0}(\vec{q}) =
\omega_N \, \rho_{N0}(\vec{q})$.

Expanding the squares of matrix elements and using the two spin-summed
relationships
$$
\sum_{spins} \rho^*_{N0}(\vec{q})J^{\alpha}_{N0}(\vec{q}) =
\frac{\vec{q}}{\vec{q}^2} \sum_{spins} | \rho_{N0}(\vec{q})|^2 \ ,\eqno({\rm
A}2)
$$
$$
\sum_{spins} J_{N0}^{\alpha}(\vec{q})J_{N0}^{\beta *}(\vec{q}) =
\sum_{spins}\left(\frac{\omega^2_N}{\vec{q}^2} | \rho_{N0}(\vec{q})|^2
\delta^{\alpha \beta} + | \vec{J}_{N0}(\vec{q}) |^2 -
\frac{3\omega^2_N}{\vec{q}} | \rho_{N0}(\vec{q}) |^2\right)
$$
$$
\times \frac{(\delta^{\alpha \beta} - {\hat q}^{\alpha} {\hat q}^{\beta})}{2}\
,
\eqno({\rm A}3)
$$
leads to the (spin-summed) result:
$$
|\rho_{N0} - \vec{\beta} \cdot \vec{J}_{N0}|^2 = |\rho_{N0}(\vec{q})|^2
\frac{(\vec{q}^2 - \omega^2_N)^2}{\vec{q}^4} + \frac{1}{2} \left(\beta^2 -
\frac{\omega^2_N}{\vec{q}^2}\right)
$$
$$
\left[ | \vec{J}_{N0}(\vec{q})|^2 -
\frac{\omega^2_N}{\vec{q^2}}|\rho_{N0}(\vec{q})|^2\right]\, . \eqno({\rm A}4)
$$
Note that the factor which multiplies (the first) $|\rho|^2$ cancels the photon
propagator and substitutes the Coulomb one.  Moreover, the quantity in
square brackets is the \underline{transverse} current (i.e., the full current
minus the longitudinal part).  This leads to the Coulomb gauge result:
\begin{eqnarray*}
\sigma_{\rm WW} = \frac{4 \pi Z_p^2  \alpha^2}{\beta^2} \sum_{N \neq 0}
\int^{\infty}_{\vec{q}^2_{min}} d \vec{q}^2  F^2_p (\vec{q}^2)
\left\{\frac{|\rho_{N0}(\vec{q})|^2}{\vec{q}^4} +
\frac{\frac{1}{2}\left(\beta^2 -
\frac{\omega^2_N}{\vec{q}^2}\right)}{(\vec{q}^2 - \omega^2_N)^2} \right.
\nonumber \\
\left. \left[|\vec{J}_{N0}(\vec{q})|^2 - \frac{\omega^2_N}{\vec{q}^2}|
\rho_{N0}(\vec{q})|^2\right]\right\}.
\end{eqnarray*}\vspace*{-0.45in}
$$
\hspace*{5.65in}(A5)
$$
Using $\beta^2 = -\frac{1}{\gamma^2} + 1$ this can be rearranged into a
form commensurate with eq. (14), after we perform the $\vec{q}^2$-integrals.
We use
$$
J^{\mu}_{N0} (\vec{q}) = \int d^3x \, e^{i\vec{q} \cdot \vec{x}} <
N|\hat{J}^{\mu}(\vec{x})|0>\, \eqno({\rm A}6)
$$
and the fact that there is no overall dependence on $\hat{q}$ after spin
sums are performed, which leads to a slightly different form of eq. (14):
\begin{eqnarray*}
\sigma_{\rm WW} = \frac{4\pi Z_p^2 \alpha^2}{\beta^2} \sum_{N \neq 0}
\frac{1}{\omega^2_N} \int d^3x \, \int
d^3{x^{\prime}}\left\{\rho_{N0}(\vec{x})
\rho^*_{N0}(\vec{x}^{\prime})\left[\frac{I_2(y)}{\gamma^2} +
I_3(y)\right.\right] \nonumber \\
\left. - \left[\vec{J}_{N0}(\vec{x}) \cdot \vec{J}^*_{N0}(\vec{x}^{\prime})
-
\rho_{N0}(\vec{x})\rho^*_{N0}(\vec{x}^\prime)\right]\left[\frac{I_1(y)}
{\gamma^2} + I_2(y)\right]\right\}.
\end{eqnarray*}\vspace*{-0.45in}
$$
\nopagebreak
\hspace*{5.65in}(A7)\\
$$
One projectile form factor has been folded into each charge and current
matrix element. That is, $\rho_{N0}(\vec{x})$ is actually the convolution
of a projectile (elastic) charge density with a target (transition) density.
We have also used $y = \omega_N z = \omega_N |\vec{x} - \vec{x}^{\prime}|$.
This compact result in configuration space is finite, \underline{requires}
no cutoffs, and is relatively simple to work with.

The corresponding expression for the cross section for the  absorption of a
photon with energy, $\omega$, can be developed in the same form:
\begin{eqnarray*}
\sigma_{\gamma}(\omega) = 2 \pi^2 \alpha \sum_{N \neq 0}
\frac{\delta (\omega -\omega_N)}{\omega_N}
\int d^3x \, \int d^3{x^{\prime}}\, j_0 (y)\,
\left[\vec{J}_{N0}(\vec{x}) \cdot \vec{J}^*_{N0}(\vec{x}^{\prime})
-\rho_{N0}(\vec{x})\rho^*_{N0}(\vec{x}^\prime)\right].
\end{eqnarray*}\vspace*{-0.45in}
$$
\hspace*{5.65in}(A8)\\
$$
Various sum rules can be constructed from this by integration:
$$
\sigma_{n} \equiv \int  d\omega\, \omega^n \sigma_{\gamma}(\omega )\ .
\eqno ({\rm A}9)
$$

In Appendix B the leading-order contributions are worked out in detail,
including an analytic evaluation of the cutoff, $R_\ell$(for $\ell =1$,
in the context of a simple nuclear model.

\pagebreak
\noindent{\bf Appendix B}

Complete expressions including formulae for the cutoffs ${R_\ell}$ are
difficult
to develop. They will also differ from multipole to multipole.
Nevertheless,
because most of the photodisintegration cross section at low energies
is
unretarded electric dipole in nature, it is useful to investigate how
the
cutoff arises and estimate its size according to models of the
electric-dipole transition density.

If one expands to lowest order in $z$ (viz., $(z)^0$) the factors
multiplying the current matrix elements
[${\vec{J}}_{N0}({\vec{x}})~\cdot~{\vec{J}}_{N0}^*({\vec{x}^{\prime}})$] in
eq.~(A7),
only unretarded electric-dipole transitions can result.
Using Siegert's theorem\cite{1a} to express the current in terms of the
dipole operator, $\vec{D}$, and the excitation energy, $\omega_N$,
$$
\int  d^3 x {\vec{J}}_{N0}({\vec{x}}) = i\omega_N {\vec{D}}_{N0}\ , \eqno ({\rm
B}1)
$$
one finds that the current terms contribute to the total
disintegration cross section
$$
\sigma_{\rm WW} = \frac{4 \pi Z_p^2 \alpha^2}{\beta^2} \sum_{N \neq 0}  T_{N0}\
,
\eqno ({\rm B}2)
$$
an amount
$$
T^{(J)}_{N0} = |{\vec{D}}_{N0}|^2 (\ln(\gamma)-{\beta^2 \over 2})\ . \eqno
({\rm B}3)
$$
Note that these terms neither require nor generate a cutoff in the logarithm.

The corresponding constant factors multiplying
$\rho_{N0}({\vec{x}}) \rho_{N0}^*({\vec{x}}^{\prime})$
give no net contribution to this order, since
$$
\int  d^3x\, \rho_{N0}({\vec{x}}) = Q_{N0} \equiv 0\ , \eqno ({\rm B}4)
$$
because we have agreed not to include elastic scattering ($N \neq 0$).  The
first-order terms in $z^2$ generate elastic scattering plus electric-dipole
transitions:
$$
T^{(\rho )}_{N0} = \frac{1}{3} \int d^3x \int d^3{x^{\prime}}
\rho_{N0}({\vec{x}})
\rho_{N0}^*({\vec{x}^{\prime}})\, z^2 \ln(z)
$$
$$
-\frac{2}{3}|{\vec{D}}_{N0}|^2 \left(\ln\left(\frac{\omega_N}{\beta}\right) +
\frac{1}{2} \ln(\gamma) + \gamma_E -\frac{11}{6}   -
\frac{\beta^2}{4}\right)\ , \eqno ({\rm B}5)
$$
where $\gamma_E$ is Euler's constant,
${\vec{z}} = {\vec{x}} - {\vec{x}^{\prime}}$, and $z^2 = (x^2 +
{x^{\prime}}^{2}) - 2 {\vec{x}} \cdot {\vec{x}^{\prime}}$ has been used.
Combining the charge and current contributions one obtains
$$
T_{N0} = \frac{1}{3} \int d^3x \int d^3x^{\prime} \rho_{N0} ({\vec{x}})
\rho_{N0}^*({\vec{x}^{\prime}})\, z^2 \ln(z)
$$

$$
+\frac{2}{3} |{\vec{D}}_{N0}|^2 \left(\ln\left[\frac{\beta  \gamma}
{\omega_N}\right] - \gamma_E + \frac{11}{6} -
\frac{\beta^2}{2}\right)\ . \eqno ({\rm B}6)
$$
This can be rearranged into the conventional Weiz\"acker-Williams\cite{2a}
form
$$
T_{N0} = \frac{2}{3} |{\vec{D}}_{N0}|^2 \left(\ln\left(\frac{2 \beta
\gamma}{\omega_N R_1}\right)- \gamma_E -
\frac{\beta^2}{2}\right)\ , \eqno ({\rm B}7)
$$
if we define:
$$
\ln(R_1/2) = -\frac{11}{6} - \frac{\int d^3x \int
d^3x^{\prime} \rho_{N0}({\vec{x}}) \rho_{N0}^*({\vec{x}^{\prime}})\,
z^2  \ln(z)}
{2|{\vec{D}}_{N0}|^2}\ . \eqno ({\rm B}8)
$$
Given any (electric-dipole)
transition density, the p-wave part of $z^2 \ln(z)$ can easily be
projected out and the double integral performed. The cutoff,$R_1$, comes
from the $I_3$-term in eq.~(A7), which itself comes solely from
Coulomb scattering (i.e., the first term in eq.~(A5)).

If the integration variables in eq.~(B8) are changed to ${\vec{z}}$ and
${\vec{x}^\prime}$, we find
$$
\int d^3x \int d^3x^{\prime} \, \rho_{N0} ({\vec{x}}) \rho_{N0}^*
({\vec{x}^{\prime}})\, z^2 \ln(z)
$$
$$
= \int d^3z\, F_N ({\vec{z}})\, z^2 \ln(z)\ , \eqno ({\rm B}9)
$$
where
$$
F_N ({\vec{z}}) = \int d^3x^{\prime}\, \rho_{N0} ({\vec{z}} +
{\vec{x}}^{\prime})
\rho_{N0}^* ({\vec{x}}^{\prime})\ , \eqno ({\rm B}10)
$$
which is a convolution of transition densities. Elastic densities of this
type have appeared in atomic calculations\cite{3a,4a}.
This form is particularly convenient for deriving
asymptotic forms, as well. The drawback is the required construction
of the
second-order (convoluted) transition density. We note that the
logarithm in eq.~(B8) is the only one contained
in $I_{3}(y)$, but it contributes to  all Coulomb multipoles.

A complete model of the electric-dipole transition density is needed in order
to
construct $F_N ({\vec{z}})$. One such model is the Goldhaber-Teller
model\cite{5a}, which
we sketch below. The transition charge operator $\rho_{N0}(\vec{x})$ can be
represented by the surface-peaked function
$$
\rho_{N0}(\vec{x}) \equiv \langle Nm | \hat{\rho} ({\vec{x}}) |0\rangle
= -\lambda_{N0}\, \hat{{\epsilon}}_m^*
\cdot {\vec{\nabla}} \rho_0(x) \ ,  \eqno ({\rm B}11)
$$
where $\langle 0|\hat{\rho}({\vec{x}})|0\rangle  \equiv \rho_0(x)$ is the
ground-state charge density, from which it follows that
$$
\int d^3x \, {\vec{x}}\, \rho_{N0}({\vec{x}}) = \lambda_{N0}\,
\hat{{\epsilon}}^*_m
\equiv {\vec{D}}_{N0}\ . \eqno ({\rm B}12)
$$
We have defined $\int d^3x \, \rho_0 (x) = 1$ and used the spherical
projection
operators $\hat{{\epsilon}}_m^*$ and the Wigner-Eckart\cite{6a} theorem to
determine the
dependence of the transition matrix element on the magnetic quantum
number ($m$) of
the final (dipole) state. For this model then, $|{\vec{D}}_{N0}|^2 =
|\lambda_{N0}|^2$ and the last term in eq.~(B8) can be rewritten as
$$
\int d^3x\, \rho_0 (x) \int d^3x^{\prime} \rho_0 (x^{\prime})
(\ln(z) + \frac{5}{6})
$$
$$
= \int  d^3z\, \rho_{(2)} (z) (\ln(z) + \frac{5}{6})    \eqno ({\rm B}13)
$$
$$
= \frac{1}{3}
+ \frac{1}{4} \int d^3x \rho_0(x) \int  d^3x^{\prime} \rho_0(x^{\prime})
\, [(x + x^{\prime})^2 \ln(x +  x^{\prime})
- (x - x^{\prime})^2 \ln | x - x^{\prime}|] /x x^{\prime} \ ,
$$
where the elastic counterpart of $F_N ({\vec{z}})$ is the Zemach
density\cite{3a}:
$$
\rho_{(2)} (z) = \int d^3x^{\prime}\, \rho_0(|{\vec{z}} + {\vec{x}}^{\prime}|)
\rho_0 (|{\vec{x}}^{\prime}|)\ . \eqno ({\rm B}14)
$$
The latter density plays a critical role in the nuclear-size
modification of
the hyperfine splitting\cite{3a} and Lamb shift\cite{4a} in atoms. Note that if
$\rho_0$ is
properly normalized, then $\rho_{(2)}$ is also. The final form
in eq.~(B13) results from
performing the average over the angle between $\hat{x}$ and
$\hat{x}^{\prime}$
contained in ${\vec{z}}$. This collective model then leads to the very
simple
result:
$$
\ln(R_1/2) = - 1 + \int d^3 z\, \rho_{(2)}(z)\, \ln(z)\ .
\eqno ({\rm B}15)
$$

To go further requires another model assumption. For simplicity we
choose a
uniform (liquid-drop) charge density:  $\rho_0 (r) = \frac{3 \theta
(R_t - r)}{4 \pi R_t^3}$, which leads to
$$
\rho_{(2)} (r) = \frac{3}{4 \pi R_t^3}\, \theta (2R_t - r) \left( 1 -
\frac{3r}{4R_t} + \frac{r^3}{16R_t^3} \right)\ . \eqno ({\rm B}16)
$$
The integral in eq.~(B15) has the value $\ln(2R_t) -\frac{3}{4}$, which
leads
finally to
$$
\ln(R_1/2) = \ln(2 R_t) - \frac{7}{4}\ , \eqno ({\rm B}17)
$$
or $\frac{R_1}{R_t} = 0.695$, where $R_t = 1.2~A_t^{\frac{1}{3}}$ fm.

This derivation ignores one important piece of physics. The
projectile, as well
as the target, has a finite size. Because in momentum space the form
factors of
initial and final projectile and target are all multiplied together,
the
projectile charge density will enter as a convolution with the
(transition)
density of the target. For example, folding the target density
in the previous model (with radius $R_t$) with a projectile whose
density is $\rho^{\prime}_0$ (with radius $R_p$) leads to a
transition density
$$
\hat{\rho} ({\vec{x}}) = \int d^3\, x^{\prime}\, \rho^{\prime}_0 (|{\vec{x}}
- {\vec{x}}^{\prime}|)\, \hat{{\epsilon}}^*_m \cdot {\vec{\nabla}}^{\prime}
\rho_0 (x^{\prime})
= \hat{{\epsilon}}^*_m \cdot {\vec{\nabla}} \bar{\rho}_{(2)} (x)\ ,
\eqno  ({\rm B}18)
$$
where $\bar{\rho}_{(2)}$ is obtained by folding $\rho_0$ and
$\rho^{\prime}_0$ together.
Thus, we can use eq.~(B15) if we fold two densities
($\bar{\rho}_{(2)}$)
together. This density, $\rho_{(4)}(x)$, can be shown to be
independent of the
order in which the folding is performed. Consequently, one can form
$\rho_{(4)}$ by
first folding two densities $\rho_0$ together, then
separately folding together two  densities
$\rho^{\prime}_0$, and finally folding these two (folded) densities together.

The resulting forms are rather complicated, and we simply state the result:
$$
\langle \ln (z)\rangle  = \int d^3z\, \rho_{(4)}(z)\, \ln(z)
= \ln(2(R_t + R_p))
$$
$$
+ (R_p - R_t)^8 \frac{(R_p + 2R_t)(R_t + 2R_p)(2R_p^2 + 11R_t R_p +
2R_t^2)}{2100 R_t^6 R_p^6} \ln\left| \frac{R_t - R_p}{R_t + R_p}\right|
$$
$$
-R_p^4 \frac{(4R_p^2 - 81R_t^2)}{1050 R_t^6} \ln\left(\frac{R_p}{R_t + R_p}
\right) - R_t^4 \frac{(4R_t^2 - 81R_p^2)}{1050 R_p^6} \ln
\left(\frac{R_t}{R_t + R_p}\right)
$$
$$
+ \frac{R_p^4}{525 R_t^4} + \frac{R_t^4}{525 R_p^4} +
\frac{107R_t^2}{700 R_p^2} + \frac{107R_p^2}{700R_t^2} -
\frac{16661}{12600}\ . \eqno ({\rm B}19)
$$
There are two simple limits:\\
$$
lim_{R_p \rightarrow 0} \left[ \langle \ln(z)\rangle
- \ln(2(R_t + R_p)) \right] =  -\frac{3}{4}
$$
and
$$
lim_{R_p \rightarrow R_t} \left[ \langle \ln(z)\rangle
- \ln(2(R_t + R_p)) \right]
= -\frac{11}{75} \ln(2) -\frac{1823}{1800}\ . \eqno ({\rm B}20)
$$
The terms in eq.~(B19) which supplement $\ln(2(R_t+R_p))$
(which we denote by $s(\frac{R_p}{R_t})$) vary monotonically from -0.75
to -1.114 as $R_p$ is varied from $0$ to $R_t$.  Thus we can rewrite
eq.~(B15) as
$$
\ln(R_1/2) = \ln(2(R_t + R_p)) - 1 + s\ , \eqno ({\rm B}21)
$$
or
$$
R_1 =  \delta \, (R_t + R_p)\ ,      \eqno ({\rm B}22)
$$
where $R_{p,t} = 1.2\, A_{p,t}^{\frac{1}{3}}$\, , and $\delta =
4 e^{ - 1 + s}$ varies from 0.70 to 0.48 for the range of
variation of $R_p$ above. Previously, a value of 1.0 was
recommended\cite{7a}.

Although the model used here can be criticized on the basis that it
is too simple to be realistic, our derivation nevertheless represents
an exact solution to the
problem for a simple collective model of electric-dipole transitions
(Goldhaber-Teller model with a uniform charge distribution) in the
unretarded dipole
limit. The troublesome cutoff is determined by details of the
electric-dipole transition charge density.

Finally, we can express our results in a particularly simple and elegant
form which emphasizes well-known photonuclear sum rules. We find
$$
\sigma_{\rm WW} = \frac{2 \alpha\, \sigma_{-1} Z_p^2}{\pi \beta^2}
\left(\ln\left(\frac{2 \beta \gamma}{\bar{\omega} R_1}\right)- \gamma_E -
\frac{\beta^2}{2}\right)\ , \eqno ({\rm B}23)
$$
where $\sigma_{-1}$ is defined in eq.~(A9) and has the experimental
value, 0.22(2)$\, A_t^{\frac{4}{3}}$~mb for medium to heavy nuclei,
and $\bar{\omega}$ is the mean photon energy, which can be taken to
be the giant resonance energy, 79$\, A_t^{-\frac{1}{3}}$~MeV\cite{6b}.
The factors in front of the bracket are (numerically) consistent with
$Z_p^2  A_t^{\frac{4}{3}}/\beta^2 \, \mu$b. The argument of the
logarithmn is $\xi\beta\gamma$, where $\xi$ varies from 6.0 to 4.3
as $R_p$ varies from 0 to $R_t$.

\pagebreak

\pagebreak
\centerline{Figure Captions}
\begin{itemize}
\item{} Figure 1 Electromagnetic excitation process.
\item{} Figure 2 Ratio of the quantum to classical excitation cross
sections for $^{12}$C on $^{197}$Au as a function of projectile energy.
\item{} Figure 3 Ratio of the quantum to semiclassical excitation cross
sections for $^{197}$Au on $^{197}$Au as a function of projectile energy.
\item{} Figure 4 Ratio of the quantum to classical excitation cross
sections for $^{12}$C on $^{197}$Au as a function of transition energy.
\item{} Figure 5 Ratio of the quantum to classical excitation cross
sections for $^{12}$C on $^{197}$Au as a function of transition energy.
\end{itemize}
\vfill\eject

\begin{tabbing}
xxxxxxxxx\=xxxxxxx\=xxxxxxx\=xxxxxxxxxx\=xxxxxxxxxxxxxxxxx\=xxxxxxxxxxxxxxxxx
xx\=xxxxxxxxxxxxxxxxx\= \kill
\> \> {$\sigma^{class}$(mb)}\> \> {$\sigma^{quantum}$(mb)} \>
{$\sigma^{EXPT}$(mb)}\\
\underline{Proj} \>
\underline{$\gamma$} \> \underline{E1 \hspace*{0.45in} E2}\> \> \underline{E1
\hspace*{0.45in} E2} \> \underline{\hspace*{.90in}} \\
\\
$^{12}$C \>  3.23 \> \hspace*{0.01in} \hspace*{0.10in} 42
\hspace*{0.35in} 9 \>\> \hspace*{0.15in} 36 \hspace*{0.35in} 4 \>
\hspace*{0.12in} 75 $\pm$ \hspace*{0.05in} 14 \hspace*{0.10in} \\
\\
$^{20}$Ne \> 3.23 \> \hspace*{0.09in} 111 \hspace*{0.28in} 22 \>\>
\hspace*{0.15in} 95 \hspace*{0.35in} 9 \> \hspace*{0.05in} 153 $\pm$
\hspace*{0.05in} 18 \\
\\
$^{40}$Ar \> 2.91 \> \hspace*{0.07in} 315 \hspace*{0.28in} 62 \>\>
\hspace*{0.05in} 262 \hspace*{0.30in} 25 \> \hspace*{0.05in} 348 $\pm$
\hspace*{0.05in} 34 \\
\\
$^{56}$Fe \> 2.81 \> \hspace*{0.07in} 614 \hspace*{0.20in} 120 \> \>
\hspace*{0.05in} 506 \hspace*{0.30in} 47 \> \hspace*{0.05in} 601 $\pm$
\hspace*{0.05in} 54 \\
\\
$^{139}$La \> 2.34 \> 2,190 \hspace*{0.20in} 462 \>\> 1,738
\hspace*{0.20in} 171 \>  1,970 $\pm$ 130 \\
\\
$^{238}$U \> 2.0 \> 4,337 \hspace*{0.10in}
1,045 \>\> 3,388 \hspace*{0.20in} 365 \>  3,160 $\pm$ 230 \\

\end{tabbing}

\noindent Table 1 Predictions of Weissacker-Williams-Fermi and quantum
theory for single neutron removal from $^{197}$Au using the
Goldhaber-Teller model for the transition densities. Data from references
2 and 12.

\end{document}